\documentclass[runningheads]{llncs}

\usepackage{graphicx}
\usepackage{multirow}

\begin{document}
\title{Semantics Guided Disentangled GAN for Chest X-ray Image Rib Segmentation\thanks{Supported by the National Natural Science Foundation of China (62106006) and the Natural Science Foundation of Anhui Higher Education Institution (KJ2020A0040). }}

\author{Lili Huang\inst{1}\and
Dexin Ma \inst{1}\and
Xiaowei Zhao\inst{2}\and
Chenglong Li\inst{2,3}\and
Haifeng Zhao\inst{1}\and
Jin Tang\inst{1} \and
Chuanfu Li\inst{3}
}

\authorrunning{L. Huang et al.}
%
\institute{School of Computer Science and Technology, Anhui University, Hefei 230601, China \and
School of Artificial Intelligence, Anhui University, Hefei 230601, China \and
Anhui Provincial Key Laboratory of Security Artificial Intelligence, Anhui University, Hefei 230601, China \and
The First Affiliated Hospital, Anhui University of Chinese Medicine, Hefei 230031, China \\
}
\maketitle              
\begin{abstract}
The label annotations for chest X-ray image rib segmentation are time consuming and laborious, and the labeling quality heavily relies on medical knowledge of annotators. To reduce the dependency on annotated data, existing works often utilize generative adversarial network (GAN) to generate training data. However, GAN-based methods overlook the nuanced information specific to individual organs, which degrades the generation quality of chest X-ray image. Hence, we propose a novel Semantics guided Disentangled GAN (SD-GAN), which can generate the high-quality training data by fully utilizing the semantic information of different organs, for chest X-ray image rib segmentation. 
In particular, we use three ResNet50 branches to disentangle features of different organs, then use a decoder to combine features and generate corresponding images. 
To ensure that the generated images correspond to the input organ labels in semantics tags, we employ a semantics guidance module to perform semantic guidance on the generated images. 
To evaluate the efficacy of SD-GAN in generating high-quality samples, we introduce modified TransUNet(MTUNet), a specialized segmentation network designed for multi-scale contextual information extracting and multi-branch decoding, effectively tackling the challenge of organ overlap. We also propose a new chest X-ray image dataset (CXRS). It includes 1250 samples from various medical institutions. Lungs, clavicles, and 24 ribs are simultaneously annotated on each chest X-ray image. The visualization and quantitative results demonstrate the efficacy of SD-GAN in generating high-quality chest X-ray image-mask pairs. Using generated data, our trained MTUNet overcomes the limitations of the data scale and outperforms other segmentation networks.

\end{abstract}

\section{Introduction}
Chest X-ray image showcases a variety of anatomical structures, such as ribs, lungs and clavicles. The shape and position information of the ribs are crucial indicators for evaluating pulmonary diseases. Chest X-ray image rib segmentation plays a crucial role in computer-aided diagnostic systems for detecting lung pathologies and rib fractures, significantly aiding in the quantification of diseases. Several deep learning networks have been proposed for the automated chest X-ray image rib segmentation\cite{liu2019automatic}\cite{wang2020mdu}\cite{wang2021rib}\cite{SINGH2022104831}\cite{wessel2019sequential}.
\begin{figure}[!t]
\centerline{\includegraphics[width=\columnwidth]{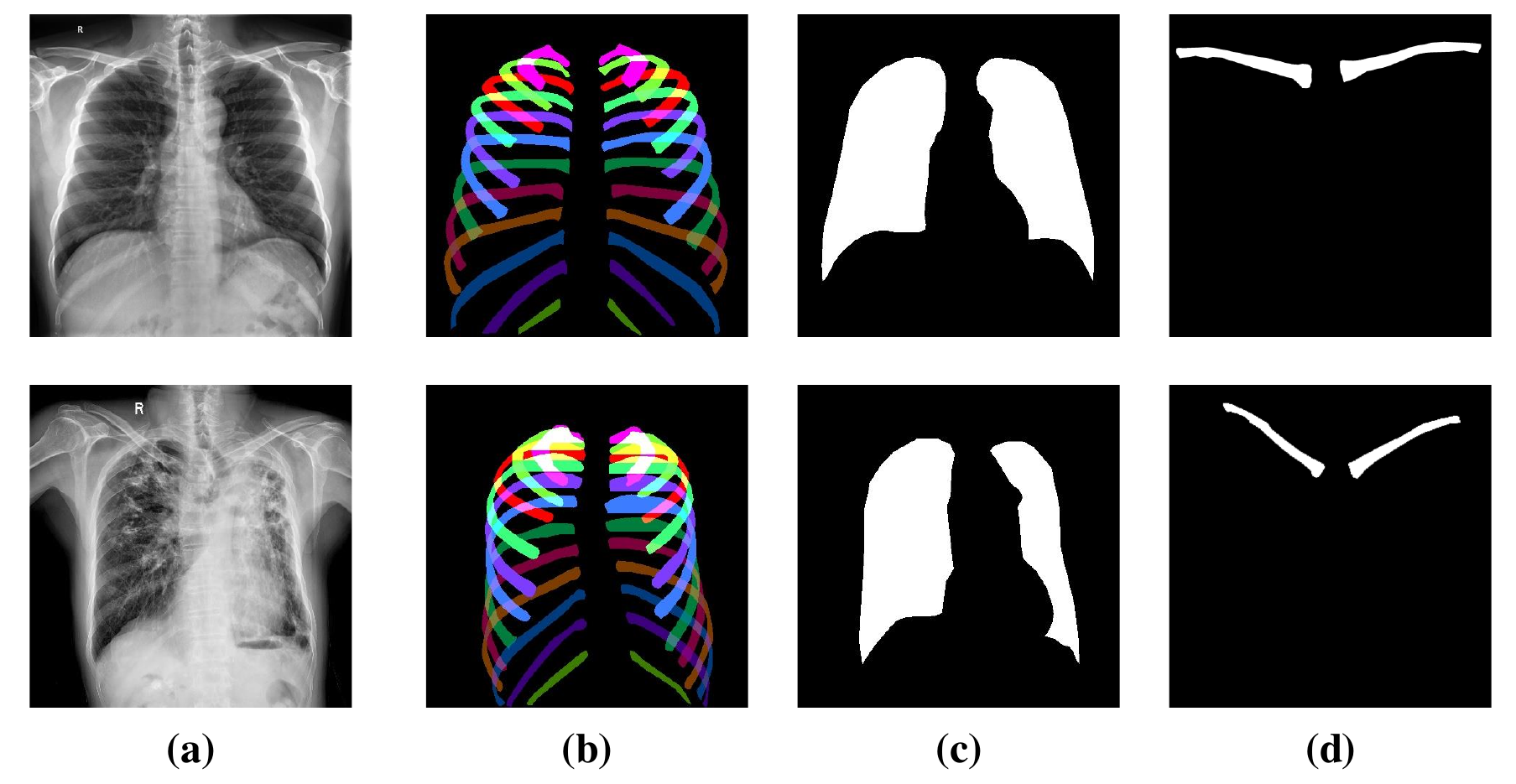}}
\caption{\textbf{The samples in CXRS.} The upper row depicts healthy samples. The lower row represents samples from patients. (a) Original image, (b) Ribs mask. (c) Lungs mask. (d) Clavicles mask. Different color represents different rib. The same color is used for symmetric ribs on the left and right sides.} 
\label{fig1}
\end{figure}

Constrained by the scale of the dataset, the more effective transformer-based segmentation network developed on natural images cannot be applied to chest X-ray image with good results. Annotating such a medical dataset including 24 ribs simultaneously is an expert-oriented, expensive, and time-consuming task. Firstly, lung markings, trachea, and blood vessels in the chest X-ray image lead to poor image contrast and limited edge definition of the ribs. Secondly, various lung diseases, such as calcified lesions, can cause abnormal local gray values in the chest X-ray image. Moreover, the presence of overlapping rib regions negatively affects the segmentation accuracy. 

Researchers have turned to traditional image augmentation techniques and generative adversarial network (GAN) to bolster the number of training samples. However, these traditional methods often fail to accurately represent the complex and varied attributes of lesions in chest X-ray image. Additionally, earlier augmentation networks based on GAN still need to maintain stringent semantic coherence between the generated images and their associated mask pairs, thus limiting the potential for performance improvement.

To solve the above problems, we present the Semantics guided Disentangled GAN (SD-GAN) for image augmentation. The network includes a generator, a discriminator, and a semantics guidance module. The generator uses three ResNet50 branches to disentangle features of ribs, lungs and clavicles. Then a feature decoder is used to combine features and generate corresponding chest X-ray image with the same resolution as the input masks. In order to make the generated samples more realistic,the discriminator is responsible for discerning real images from fake ones. To ensure that the generated images correspond to the input organ labels in semantics tags, we employ a semantics guidance module to perform semantic guidance. Specifically, we use UNet to segment ribs, lungs, and clavicles from synthetic chest X-ray image. We use  binary cross-entropy loss (BCELoss) to constrain the maps of semantic segmentation to be consistent with the input organ masks. 

The synthetic chest X-ray image, along with the input masks of different organs, form a novel image-mask pair. Notably, incorporating the semantics guidance module in our method promotes greater semantic consistency in the generated image-mask pairs compared to existing networks. 

In the inference stage, we generate new masks by applying affine transformations to the original masks. Subsequently, we utilize the generator in the SD-GAN to synthesize corresponding chest X-ray image. The synthetic data will be used to train a more robust rib segmentation network along with the original data.
\begin{figure*}
\includegraphics[width=\textwidth]{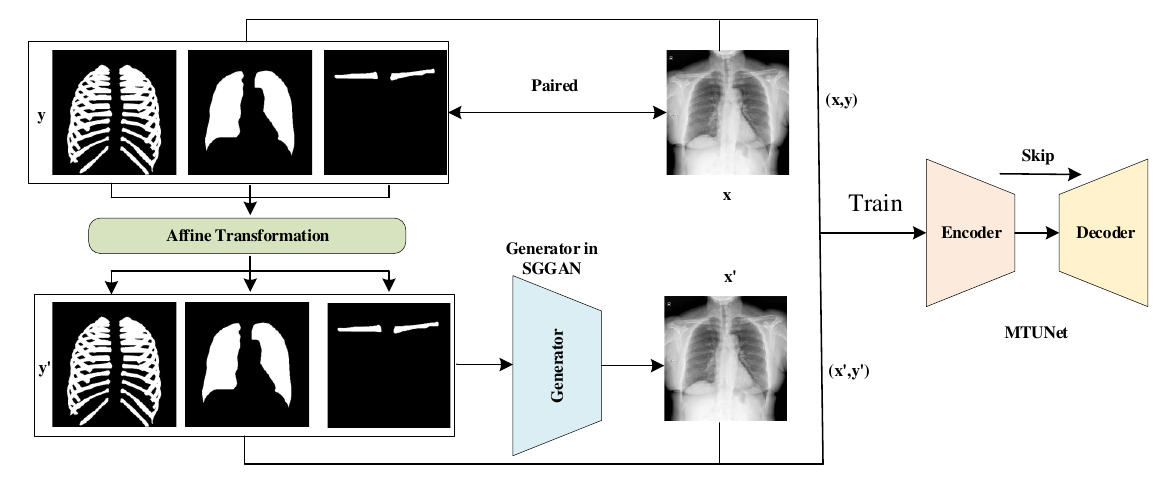}
\caption{\textbf{Overview of training framework named SGTUNet.} Given a chest X-ray image $x$ and its corresponding labels $y$, SD-GAN can generate a new paired sample \{$x'$,$y'$\}. The $y'$ is obtained by performing affine transformations on $y$. After that, $x'$ is synthesized by feeding $y'$ into the generator in SD-GAN. The generated sample \{$x'$,$y'$\} will be fed into MTUNet together with real sample \{$x$,$y$\}.} 
\label{fig2}
\end{figure*}

To verify the effectiveness of SD-GAN in generating high-quality chest X-ray image-mask pairs, we propose a modified TransUNet (MTUNet) for chest X-ray image rib segmentation. In MTUNet, we apply the atrous spatial pyramid pooling block(ASPP) proposed in Deeplab v3 +\cite{chen2018encoder} to the original TransUNet. This module is used to aggregate multi-scale contextual information. Due to the fact that each pixel may belong to multiple rib regions, we replace a single segmentation head with multiple parallel segmentation heads after the decoder to obtain segmentation results for each organ. We use Sigmoid as the activation function instead of Softmax. 

Due to the unavailability of existing annotated datasets, we collect a new chest X-ray image dataset (CXRS) to verify that our method can solve the algorithm's excessive reliance on annotated data. We collected 1250 samples from different medical institutes under the guidance of expert doctors. 850 samples are obtained from individuals with average health conditions, ensuring high-quality images. The remaining 400 images are sourced from patients diagnosed with various lung diseases. The dataset includes healthy and unhealthy samples with annotations of lungs, ribs,and clavicles. Experimental results on CXRS confirm that SD-GAN significantly enhances the trained segmentation network's segmentation accuracy and generalization capability. Our contributions are  
\begin{itemize}
\item[$\bullet$] We propose the semantics guided disentangled GAN to generate highly accurate image-mask pairs by combining the semantic information of different organs.
\item[$\bullet$] We propose a modified TransUNet for accurate chest X-ray image rib segmentation, by applying the atrous spatial pyramid pooling block and multiple segmentation heads.
\item[$\bullet$] We propose a new chest X-ray image dataset called CXRS. Through extensive experiments on CXRS, we demonstrate that SD-GAN outperforms other image augmentation methods for chest X-ray image rib segmentation.
\end{itemize}
\section{Related Work}
\subsection{GANs in Medical Image for Image Augmentation}
Using generative adversarial network for medical image synthesis can address the large and diverse annotated dataset shortage. Frid et al.\cite{frid2018gan} utilize generative adversarial network to synthesize high-quality liver focal lesions from CT images for the liver lesion classification task. Jin et al.\cite{jin2018ct} employ a 3D generative adversarial network to generate fake lung CT artificial scans with lung nodules of various dimensions at multiple locations and achieve promising results. Wang et al.\cite{wang2021rib} use CycleGAN to learn texture changes in pneumonia images and produce different annotated chest X-ray image, while existing methods do not provide such samples. Dong et al.\cite{8310638} propose the generation of ct images based on GAN. However, these networks are designed for specific tasks. Without semantics guidance, they can not ensure that synthetic labels strict match the synthetic images in semantics when they are used to generate chest X-ray images and corresponding rib labels. And they perform image to image translation without considering the relationship among anatomical structures.
\subsection{Chest X-ray Image Rib Segmentation}
Medical image segmentation is a crucial way to assist doctors in accurately diagnosing diseases. 
The earliest application of Convolutional Neural Network (CNN) in medical image segmentation is the FCN\cite{long2015fully}. And then, UNet\cite{ronneberger2015u} and it’s variants such as UNet++\cite{zhou2018unet++}, attention UNet\cite{oktay2018attention}, UNet3+\cite{huang2020unet} are proposed for specific tasks. Although these networks perform well, they can not capture long-distance dependencies. Transformer structure\cite{vaswani2017attention} can construct global contextual information, widely used in natural language processing. Chen et al.\cite{chen2021transunet} combine Transformer and Unet for medical image segmentation. Zhang et al.\cite{zhang2021transfuse} propose the TransFuse network with a parallel CNN and Transformer feature extraction module designed. Tang et al.\cite{hatamizadeh2022swin} propose a self-supervised pre-trained segmentation network based on the Swin Transformer and propose the Swin UNETR. Zhou et al.\cite{zhou2021nnformer} propose the nnFormer network, where the Transformer and CNN are used alternately in the network. Li et al.\cite{Li2023} introduce a single-point preliminary feature extraction (SPFE) module to address the problem of inadequate segmentation capability of the teeth-gingival boundary.

To the best of our knowledge, due to the lack of publicly available dataset, there are only five algorithms based on deep neural networks are used in chest X-ray image rib segmentation. Wessel et al.\cite{wessel2019sequential} use mask R-CNN for labeling and segmentation of ribs on chest X-ray image. Liu et al.\cite{liu2019automatic} use the full convolution network to classify the chest X-ray image pixel-by-pixel to segment the posterior, anterior, and clavicles. Wang et al.\cite{wang2020mdu} propose a multi-task densely connected UNet combining DenseNet and multi-task mechanism to strengthen the feature learning process, effectively improving the rib segmentation performance. Wang et al.\cite{wang2021rib} propose a hierarchical multi-scale learning framework to separate overlapping target depths, after which the segmentation accuracy of overlapping organs in the ribs is improved. 
Singh et al.\cite{SINGH2022104831} presents a robust encoder–decoder network for semantic segmentation of bone structures on normal as well as abnormal chest X-ray image.
\section{Method}
In this section, we provide a detailed introduction to the implementation of SD-GAN and MTUNet.
\begin{figure*}
\includegraphics[width=\textwidth]{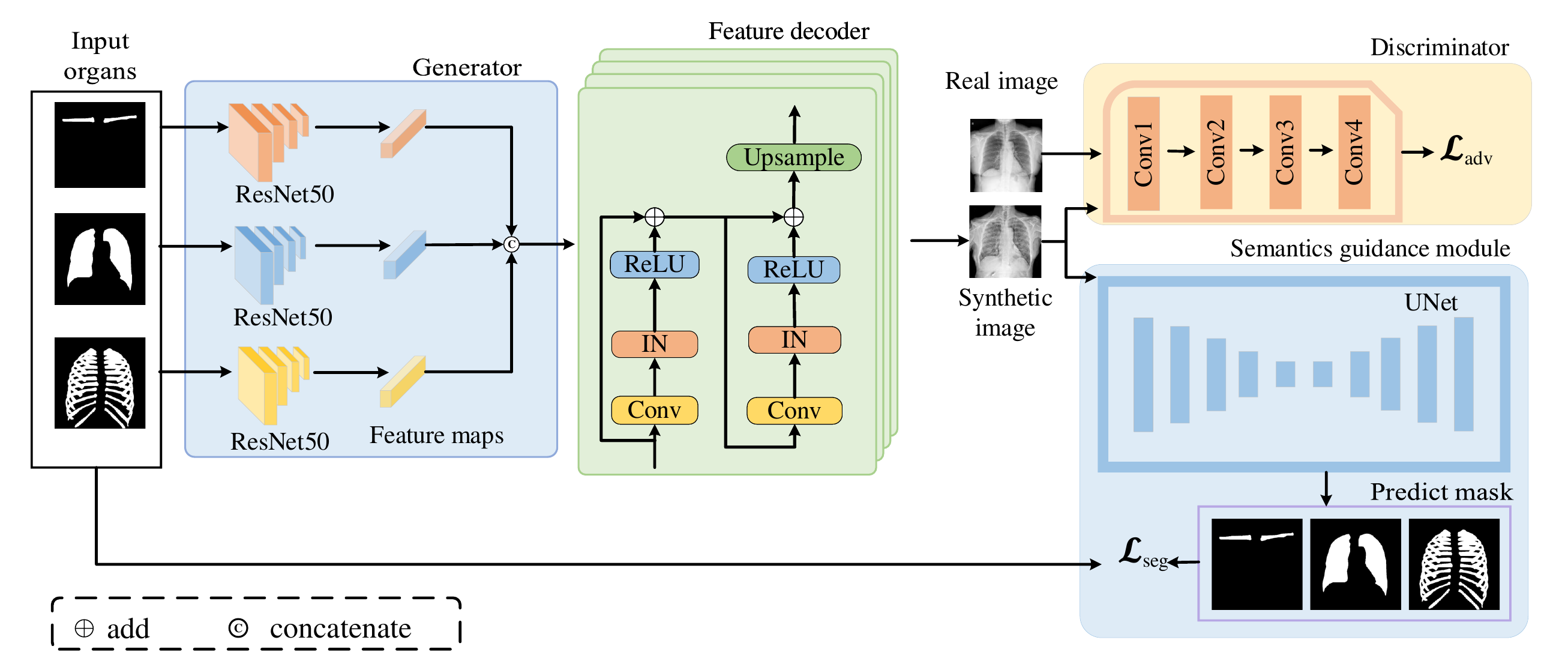}
\caption{\textbf{Overview of the proposed method, SD-GAN.} Three ResNet50 backbones are utilized to disentangle features from different organs. After concatenating the features, a synthetic image is obtained through the feature decoder. Discriminators and semantics guidance module constrain generator through $\mathcal{L}_{Adv}$ and $\mathcal{L}_{Seg}$, respectively.
}
\label{fig3}
\end{figure*}

\subsection{Semantics guided Disentangled GAN}
SD-GAN is depicted in Fig\ref{fig3}. The architecture consists of a generator, a discriminator, and a semantics guidance module. In the first stage, we employ the UNet in semantics guidance module to simultaneously segment the ribs, lungs, and clavicles in chest X-ray image using binary cross-entropy loss. In the second stage, we use a generator to learn a mapping from the input masks of different organs to corresponding images. The discriminator and the semantics guidance module are employed to supervise the generator process by combining adversarial loss and semantic segmentation loss. In the inference stage, we obtain new masks by performing affine transformations on original masks. Subsequently, we generate corresponding chest X-ray images by inputting masks of different organs into the generator. These generated image-mask pairs are used to train MTUNet and improve the segmentation accuracy. Our training framework named SGTUNet is shown in Fig\ref{fig2}. Together with real data, the masks synthesized through the affine transformation module and the images generated by the generator in SD-GAN will be used to train MTUNet. 

\subsubsection{Generator}
In the generator(G), three encoders disentangle features from different organs. These features are concatenated and used by a decoder to generate a chest X-ray image. We adopt ResNet50 on three branches with unique parameters, serving as the encoders. The ResNet50, proposed by He et al. \cite{he2016deep}, offers a balance between lightweight structure and effective feature extraction capacity, making it ideal for disentangling organ features. Specifically, $E_{1}$ disentangles features from the clavicles, $E_{2}$ disentangles features from the lungs, and $E_{3}$ disentangles features from the ribs. The decoder($D$), as depicted in Fig\ref{fig3}, generates the corresponding chest X-ray image.
The process is given as:
\begin{equation}
f_{1} = E_{1}(p_{rib}),\quad
f_{2} = E_{2}(p_{lung}),\quad
f_{3} = E_{3}(p_{clavicle}) 
\end{equation}
\begin{equation}   
x_{g} = G(p) = D(cat(f_{1},f_{2},f_{3})), \quad \
p=(p_{rib},p_{lung},p_{clavicle})
\end{equation}
Here, the $p_{rib}$, $p_{lung}$, and $p_{clavicles}$ represent the data distribution of the masks of ribs, lungs, and clavicles separately. $x_{g}$ represents the generated chest X-ray image.

\subsubsection{Discriminator}
We employ the PatchGAN as the discriminator, a framework introduced by Isola et al.\cite{isola2017image}. PatchGAN has demonstrated effectiveness in discriminating real and fake images. It penalizes the structure at the patch level, evaluating the authenticity of each N × N patch within an image. Running this discriminator across the entire image, we aggregate the responses to obtain the final output of the discriminator.
\subsubsection{Semantics guidance module}
To ensure that the generated chest X-ray image corresponds to the input organ masks in semantics tags, we employ a semantics guidance module to perform semantic guidance on the generated chest X-ray image. We utilize UNet as the semantics guidance module, commonly employed in semantic segmentation tasks. It consists of an encoder and a decoder. The encoder extracts multi-scale features while the decoder aggregates multi-scale features and generates the final output. In the first stage, we utilize all chest X-ray images and their corresponding masks to train the UNet. In the second stage, we freeze the parameters and segment lungs, ribs and clavicles form the synthetic chest X-ray image. Then we use BCELoss to constrain the maps of semantic segmentation to be consistent with the input organ masks. The loss function $\mathcal{L}$ is defined as follows:
\begin{equation}
\mathcal{L}_{GAN} = Min_{G}Max_{D}( E[D(x_{d})] + E[1-D(G(p))])
\end{equation}
\begin{equation}
\mathcal{L}_{Seg}(S(G(z)),p) = \mathcal{L}_{BCE}(S(G(z)),p) + \mathcal{L}_{Dice}(S(G(z)),p) 
\end{equation}
\begin{equation}
\mathcal{L} = \mathcal{L}_{GAN} + \mathcal{L}_{Seg}(S(G(z)),p)
)
\end{equation}
Here, $x_{d}$ represents the actual sample, $p$ represents the distribution of masks, and $S$ represents the semantics guidance module. The $\mathcal{L}_{Seg}$ combines the binary cross-entropy loss $\mathcal{L}_{BCE}$ and the dice loss $\mathcal{L}_{Dice}$ to ensure that the labels of the generated image closely match the input masks.
\subsection{Modified TransUNet for chest X-ray image rib segmentation}
\begin{figure*}[!t]
\centerline{\includegraphics[width=\textwidth]{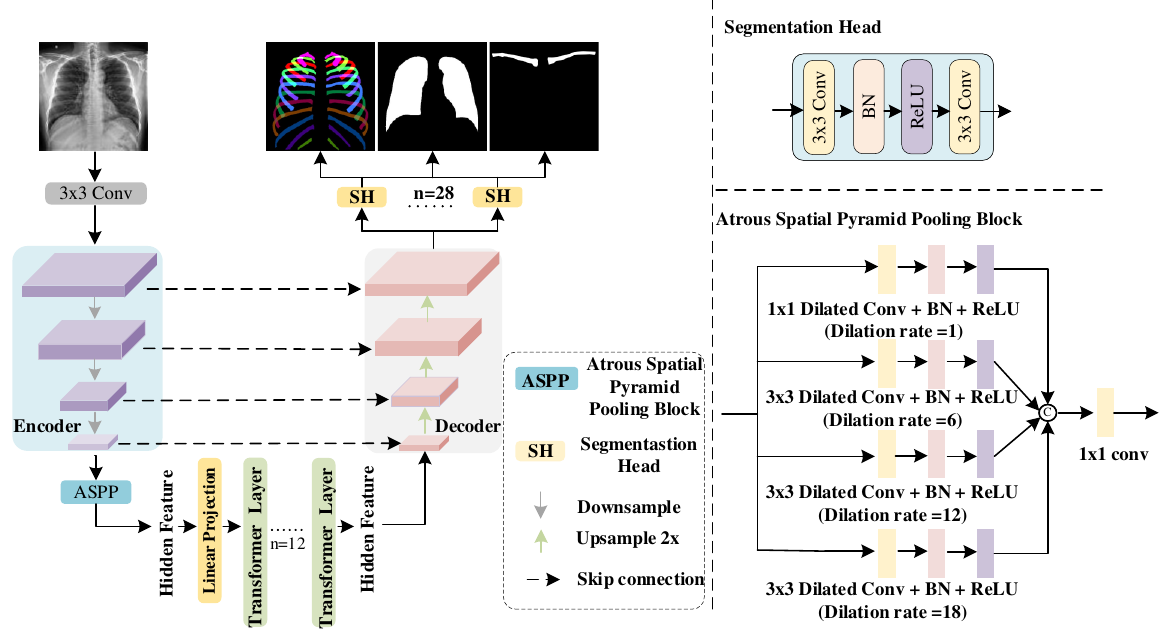}}
\caption{\textbf{Overview of MTUNet.} Based on TransUNet, the ASPP module is added after the encoder to extract multi-scale contextual features. Following the decoder, multiple segmentation heads are designed to divide each pixel into multiple organ regions.} 
\label{fig4}
\end{figure*}

In TransUNet\cite{chen2021transunet}, CNN captures local features, while the Transformer layers are responsible for modeling global contextual relationships. To further enhance its ability to extract multi-scale context information, we add the atrous spatial pyramid pooling block to its feature extractor. The overall architecture is shown in the fig\ref{fig4}. The decoder module includes four times upsampling blocks to connect low-level and high-level feature information, enabling us to retain both the edges and detail information of the organs. Due to the overlapping anatomical structures, the original TransUNet cannot segment multiple anatomical structures simultaneously. We introduce multiple segmentation heads to divide each pixel into multiple organ regions. The activation function for each segmentation head is Sigmoid instead of Softmax.
\section{Experiment}
\subsection{Dataset and metrics}
\subsubsection{Dataset}
Since the previous datasets for chest X-ray image rib segmentation are not available due to privacy concerns and in order to validate our method can overcome the label scarcity problem, we collected 1250 samples from different medical institutes under the guidance of expert doctors. 850 samples are obtained from individuals with health conditions, ensuring high-quality images. The remaining 400 images are sourced from patients diagnosed with various lung diseases. Fig\ref{fig1} showcases a selection of the collected sample data, with the upper row representing the samples from healthy individuals and the lower row displaying samples from patients.

The previous private datasets arbitrarily divide ribs into anterior and posterior ribs\cite{liu2019automatic}\cite{wang2020mdu}\cite{wang2021rib}. However, in practical clinical applications, it is crucial to determine the precise positioning of each rib accurately. In our chest X-ray image dataset(CXRS), the lungs, clavicles, and 24 ribs are annotated concurrently in each chest X-ray image. Annotating each chest X-ray image requires an expert radiologist for approximately 30 minutes. Each chest X-ray image is initially annotated by one radiologist and subsequently reviewed by another. The chest X-ray image and the corresponding annotation results are depicted in Fig\ref{fig1}. The dataset is randomly divided into training, validation, and testing sets in a 6:2:2 ratio. In the test set, 32\% of the samples are pathological, all of which are very difficult to segment.


\subsubsection{Metrics} To evaluate the segmentation performance, two standard metrics are used for evaluation, i.e., mean Intersection Over Union of different categories (mIOU) and mean Dice Similarity Coefficient of different categories (mDSC). They measure the extent of overlap between the segmentation probability map and ground truth. Higher IOU  and DSC values indicate superior segmentation performance.
\subsection{Implementation Details}
Our network is implemented in PyTorch and trained on a single NVIDIA RTX 6000 GPU. During the first stage, we utilize the Adam optimizer with an initial learning rate of 1e-4 to train the UNet. This learning rate decays to zero throughout 200 epochs. The batch size was set to 8 for this stage. In the second stage, the generator and discriminator are trained using the Adam solver with a batch size 2. All networks are trained from scratch, using a learning rate of 0.0002. For the first 100 epochs, we maintain the same learning rate and then linearly decay it to zero over the subsequent 100 epochs.

After training SD-GAN, we only use the generator to generate new image-mask pairs. The synthetic image-mask pairs are combined with the original to train MTUNet. Moreover, MTUNet is trained with the SGD optimizer, using a learning rate of 0.01, a momentum of 0.9, and a weight decay of 1e-4. The default batch size is 8. We resize all the images to a resolution of 448 × 448, along with the corresponding segmentation maps. 

\subsection{Main Results}
\begin{figure*}[h]
\centering
\includegraphics[width=\textwidth]
{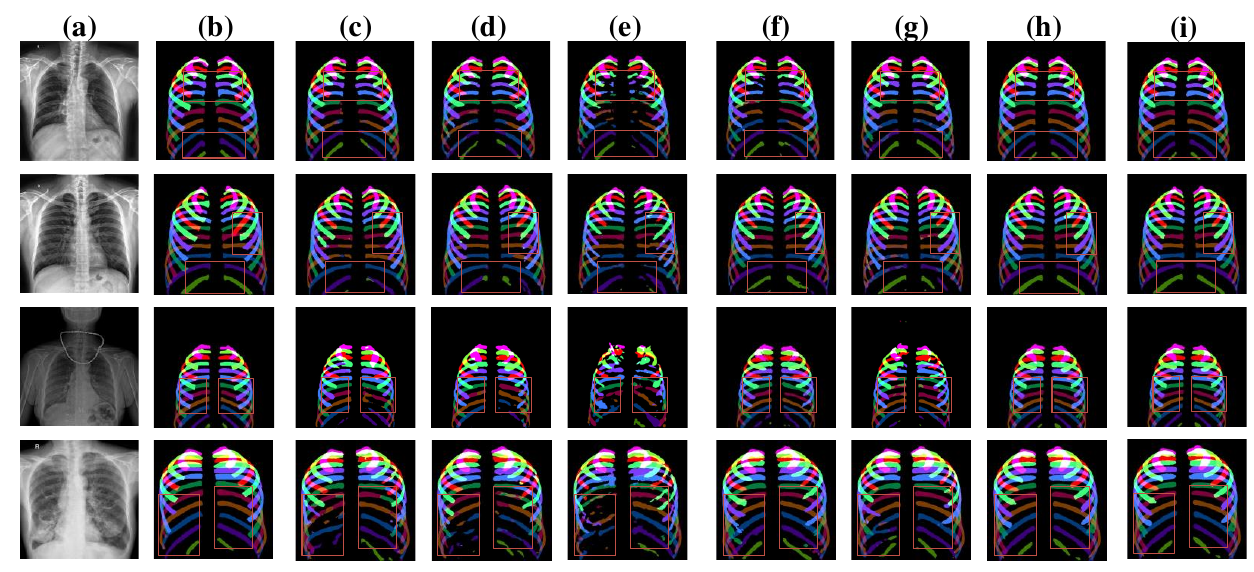}
\caption{\textbf{The result of qualitative comparison for chest X-ray image rib segmentation on CXRS by visualization.} (a) Original Chest X-ray image. (b) Ground Truth. (c) Attention UNet. (d)  UCTransNet. (e) UNext. (f) Our MTUNet augmented by traditional methods. (g) Our MTUNet augmented by CycleGAN. (h) Our MTUNet augmented by DDPM. (i) Our MTUNet augmented by SD-GAN.}
\label{fig5}
\end{figure*}
To evaluate the effectiveness of MTUNet, we compare it with widely used medical image segmentation frameworks, including UNet\cite{ronneberger2015u}, Unet++\cite{zhou2018unet++}, Attention UNet\cite{oktay2018attention}, UCTransNet\cite{wang2022uctransnet}, and UNeXt\cite{valanarasu2203unext}. To evaluate the effectiveness of the proposed SD-GAN, we conducted experiments on CXRS to generate images from masks. In order to compare and assess its performance, we compare SD-GAN with the following methods:(1)Traditional augmentation method. This approach relies on manually designed functions to deform the input image and segmentation maps simultaneously, resulting in paired deformation samples. We use transformations including rotation, translation, flipping, and scaling. However, these functions are limited to producing simple deformations and fail to simulate the various texture changes caused by different lesions present in real chest X-ray image. (2)CycleGAN\cite{zhu2017unpaired}. By employing cycle consistency loss, CycleGAN can convert images from one domain to another domain without needing paired examples. This ensures that the converted image can be accurately translated back to its original domain. (3)Denoising Diffusion Probabilistic Model(DDPM)\cite{ho2020denoising}. Utilizing masks as prompts, DDPM generates chest X-ray image from noisy images through a t-step denoising process. To comprehensively assess the impact of different image augmentation methods on segmentation performance, we conduct qualitative and quantitative comparisons of the segmentation results. 
\begin{table*}[h]
\caption{\textbf{The result of quantitative comparison between other segmentation networks and ours.}}
\centering
\label{tab1}
\begin{tabular}{c|cccccc}
\hline
\hline
\multicolumn{1}{c|}{\multirow{2}{*}{methods}} & \multicolumn{2}{c}{\makebox[0.2\textwidth]{ribs}} & \multicolumn{2}{c}{\makebox[0.2\textwidth]{lungs}} & \multicolumn{2}{c}{\makebox[0.2\textwidth]{clavicles}} \\
\multicolumn{1}{c|}{} & mIOU        & mDSC       & mIOU        & mDSC        & mIOU          & mDSC\\ \hline
UNet\cite{ronneberger2015u}                                          & 0.619                    & 0.734                    & 0.925       & 0.955       & 0.830         & 0.898         \\
UNet++\cite{zhou2018unet++}                                        & 0.683                    & 0.785                    & \textbf{0.928}       & \textbf{0.957}       & \textbf{0.848}         & \textbf{0.910}         \\
attentionUnet\cite{vaswani2017attention}                               & 0.693                    & 0.794                    & 0.927       & 0.956       & 0.845         & 0.908      \\
UCTransNet\cite{wang2022uctransnet} &0.692&0.792&0.925&0.955&0.833&0.900\\
UNeXt\cite{valanarasu2203unext}&0.572&0.699&0.919&0.951&0.757&0.852\\
MTUNet(ours)                                     & \textbf{0.751}          & \textbf{0.844}                   & 0.926       & 0.956       & 0.846         & 0.908       \\

\hline
\hline
\end{tabular}
\end{table*}

\subsubsection{Quantitative Comparison}
We perform a quantitative comparison between MTUNet and other segmentation networks. The result presented in Table~\ref{tab1} demonstrates that our method outperforms other segmentation networks. Comparison with the existing semantic segmentation methods shows that our method achieves improved segmentation results with a mIOU of 0.751, 0.926, and 0.846 for segmentation of all ribs, lungs, and clavicle, respectively. The mDSC values achieved by our network are 0.844, 0.956, and 0.908 for segmenting all ribs, lungs, and clavicles, respectively. For rib segmentation, MTUNet achieve 0.132, 0.068, 0.058, 0.059, and 0.179 improvement in the mIOU value compared with UNet, UNet++, Attention UNet, UCTransNet, and UNeXt, respectively. This can be attributed to the proficiency of SD-GAN in data synthesis and the powerful feature extraction ability of MTUNet. 

We also perform a quantitative comparison between SD-GAN and other image augmentation networks. To make a fair comparison, We use the data generated by each image augmentation method to train the MTUNet. Using SD-GAN as the image augmentation network leads to improved segmentation results with a mIOU of 0.026, 0.023, and 0.049 for rib segmentation, compared to the traditional augmentation method, CycleGAN, and DDPM, respectively. 
\subsubsection{Qualitative Comparison}
We provide the result of qualitative comparison on CXRS, as shown in Fig\ref{fig5}. It can be seen that compared with other segmentation networks, MTUNet achieves the best performance for edge location. Due to the low contrast of the 12th rib on the left and the 12th rib on the right in the image, other segmentation methods are challenging for accurate segmentation. However, due to a large number of highly matched image-mask pairs generated by SD-GAN and the powerful feature extraction ability of MTUNet, MTUNet accurately captures these organs' position and edge information. The visualization in Fig\ref{fig5}  indicates that the high-quality data generated by SD-GAN is more conducive to training a robust network than other image augmentation methods.
\begin{table*}[h]
\centering
\caption{\textbf{Quantitative comparison conducted between SD-GAN and other image augmentation methods.} The evaluation metrics are mIOU and mDSC.}
\begin{tabular}{c|cccccc}
\hline
\hline
\multicolumn{1}{c|}{\multirow{2}{*}{methods}} & \multicolumn{2}{c}{\makebox[0.2\textwidth]{ribs}} & \multicolumn{2}{c}{\makebox[0.2\textwidth]{lungs}} & \multicolumn{2}{c}{\makebox[0.2\textwidth]{clavicles}} \\
\multicolumn{1}{c|}{} & mIOU        & mDSC       & mIOU        & mDSC        & mIOU          & mDSC\\ \hline
traditional augmentation                      & 0.725                    & 0.824                    & \textbf{0.929}       & \textbf{0.957}       & 0.836         & 0.902         \\
cycleGAN\cite{zhu2017unpaired}                                      & 0.728                    & 0.828                    & 0.926       & 0.956       & 0.837         & 0.903         \\
DDPM  \cite{ho2020denoising}                                        & 0.702                    & 0.804                    & 0.928       & 0.957       & 0.833         & 0.901         \\
SD-GAN(ours)&                               \textbf{0.751}                   & \textbf{0.844}                    & 0.926       & 0.956       & \textbf{0.846}         & \textbf{0.908}        \\
\hline
\hline
\end{tabular}
\end{table*}
\subsection{Ablation Studies}
For the ablation studies, we only perform chest X-ray image rib segmentation on CXRS for all our experiments because it is the most critical challenge for existing networks.
\subsubsection{Ablation Studies on Synthetic Data Volume} We utilize affine transformation techniques on original masks to generate additional masks. Subsequently, we employ the generator in SD-GAN to synthesize the corresponding chest X-ray image. This method is employed to expand the dataset. We train MTUNet using various volumes of synthetic data to investigate the influence of synthetic data volume on the experimental results. Table\ref{tab3} presents the impact of synthetic data volume on the mIOU and mDSC of chest X-ray image rib segmentation. A significant improvement in segmentation performance is not observed until the synthetic data volume is seven times that of accurate data. 
\begin{table}[ht]
\caption{\textbf{Ablation study of the impact of synthetic data volume on mIOU and mDice during training MTUNet for chest X-ray image rib segmentation.}}
\centering
\label{tab3}
\begin{tabular}{cc|ccc}
\hline
\hline
\makebox[0.2\columnwidth]{synthetic data} & \makebox[0.2\columnwidth]{  real data } & \makebox[0.1\columnwidth]{mIOU} & \makebox[0.1\columnwidth]{mDice} \\
\hline
750 &  750   & 0.752    & 0.844 \\
1500 &  750   & 0.755   & 0.847 \\
2250 &  750    & 0.756   & 0.848 \\
3000 &  750   & 0.760   & 0.856   \\
3750 &  750   & 0.760   & 0.856   \\
4500 &  750   & 0.760   & 0.856  \\
\hline
\hline
\end{tabular}
\end{table}
\subsubsection{Ablations Studies on Proposed Modules} We use TransUNet with multiple segmentation heads as the baseline. As shown in Table\ref{tab4}, 'baseline+SD-GAN+ASPP' achieves the best performance. When there is a sufficient amount of data, the Transformer architecture fully leverages its potential to capture global contextual information.
\begin{table}
\caption{\textbf{Ablation studies on proposed modules for chest X-ray image rib segmentation.}}
\centering
\label{tab4}
\begin{tabular}{c|cc}
\hline
\hline
\makebox[0.2\textwidth]{Method}  & \makebox[0.1\textwidth]{mIOU} & \makebox[0.1\textwidth]{mDice} \\
\hline
baseline & 0.725  & 0.824 \\
baseline + ASPP&  0.732  & 0.827  \\
baseline + SD-GAN& 0.747   & 0.840 \\
baseline + SD-GAN + ASPP  & 0.752    & 0.844   \\ 
\hline
\hline
\end{tabular}
\end{table}
\section{Conclusion}
We propose a Semantics Guided Disentangled GAN, capable of synthesizing realistic chest X-ray image that maintain semantic consistency with the input anatomical structures. We synthesize a large amount of annotated data through affine transformations and the generator in SD-GAN. Moreover, we propose MTUNet for effective chest X-ray image rib segmentation. We conduct comprehensive experiments on CXRS, demonstrating the superior performance of MTUNet compared to other segmentation methods. Furthermore, we validate that SD-GAN outperforms other image augmentation methods in this task. Moving forward, we will continue exploring advanced segmentation networks for chest X-ray image rib segmentation. 
\subsubsection{Limitations:}
We synthesize many chest X-ray image-mask pairs by utilizing additional labeling data of ribs, lungs, and clavicles. However, when these data are used for segmentation, only the segmentation performance of ribs is improved, while that of the lungs and clavicles remains unchanged. Comparing the segmentation results of different segmentation networks, we believe that the segmentation performance of the lungs and clavicles has reached bottlenecks. We will explore better methods to address these issues.

\end{document}